\documentclass[aps,twocolumn,showpacs,preprintnumbers,amsmath,amssymb,superscriptaddress,prr,longbibliography]{revtex4-1}
\usepackage{graphicx}% Include figure files
\usepackage{dcolumn}% Align table columns on decimal point
\usepackage{bm}% bold math
\usepackage{indentfirst}
\usepackage{amsmath}
\usepackage{amssymb}
\usepackage{mathrsfs}
\usepackage{color}

\newcommand{\RNum}[1]{\uppercase\expandafter{\romannumeral #1\relax}}

\begin{document}

\title{Non-equilibrium dynamics of ultracold lattice bosons inside a cavity }

\author{Huan Wang}
\thanks{These authors contribute equally to this work.}
\affiliation{MOE Key Laboratory for Nonequilibrium Synthesis and Modulation of Condensed Matter,Shaanxi Province Key Laboratory of Quantum Information and Quantum Optoelectronic Devices, School of Physics, Xi'an Jiaotong University, Xi'an 710049, China}

\author{Xiayao He}
\thanks{These authors contribute equally to this work.}
\affiliation{MOE Key Laboratory for Nonequilibrium Synthesis and Modulation of Condensed Matter,Shaanxi Province Key Laboratory of Quantum Information and Quantum Optoelectronic Devices, School of Physics, Xi'an Jiaotong University, Xi'an 710049, China}

\author{Shuai Li}
\affiliation{MOE Key Laboratory for Nonequilibrium Synthesis and Modulation of Condensed Matter,Shaanxi Province Key Laboratory of Quantum Information and Quantum Optoelectronic Devices, School of Physics, Xi'an Jiaotong University, Xi'an 710049, China}

\author{Hongrong Li}
\affiliation{MOE Key Laboratory for Nonequilibrium Synthesis and Modulation of Condensed Matter,Shaanxi Province Key Laboratory of Quantum Information and Quantum Optoelectronic Devices, School of Physics, Xi'an Jiaotong University, Xi'an 710049, China}

\author{Bo Liu}
\email{liubophy@gmail.com}
\affiliation{MOE Key Laboratory for Nonequilibrium Synthesis and Modulation of Condensed Matter,Shaanxi Province Key Laboratory of Quantum Information and Quantum Optoelectronic Devices, School of Physics, Xi'an Jiaotong University, Xi'an 710049, China}

\begin{abstract}
We study the non-equilibrium quench dynamics crossing a continuous phase transition between the charge density wave (CDW) and supersolid (SS) phases of a bosonic lattice gas with cavity-mediated interactions. When changing the hopping amplitude in the Hamiltonian as a function of time, we investigate the scaling behavior of the correlation length and vortex density with respect to the quench time and find that there is a threshold of the quench rate separating two distinct scaling regimes. When slowly varying the system below that threshold, we find a power-law scaling as predicted by the Kibble-Zurek
mechanism (KZM). While considering fast quench above that threshold, a deviation from the KZM prediction occurs, manifested by a saturation of the defect density. We further show that such distinct scaling behaviors during different dynamic procedures can be understood through comparing the relaxation time and the quench rate.
\end{abstract}

\maketitle

\section{Introduction}

The study of dynamics of quantum many-body system is one of the most exciting frontiers in modern condensed matter physics~\cite{2011_Polkovnikov_etal,RevModPhys2014}. The so-called Kibble-Zurek mechanism (KZM) has been used to understand
certain universal features of the dynamics across a continuous phase transition both in classical~\cite{1976_Kibble,1980_Kibble,1985_Zurek,1993_Zurek,1996_Zurek} and quantum~\cite{2005_Damski,2005_Zurek_Dorner_zoller,2005_Polkovnikov,2005_Dziarmaga,2007_Mukherjee_etal,2008_Sen_etal,2010_Dziarmaga,2011_Polkovnikov_etal,RevModPhys2014,2009_Cincio_etal} realms. Based on conventional scaling laws near the criticality, KZM predicts a universal scaling for resulting density of defects when the system crosses a second-order phase transition by a linear quench. Recently, tremendous amount of efforts in both theoretical and experimental studies have been triggered to explore such quench dynamics, for instance in liquid crystals~\cite{Science1994,SanatanPhysRevLett1999}, quantum optical systems~\cite{XuPhysRevLett2014}, superconducting films~\cite{ManivPhysRevLett2003,GolubchikPhysRevLett2010}, trapped ions~\cite{2012_Chandran_etal,2016_Francuz_etal,2009_Bermudez_etal,2010_Bermudez_etal,2014_Dziarmaga_Zurek,2017_Gardas_etal}, as well as ultracold gases~\cite{2015_Navon_etal,2011_Chen_etal,2015_Braun_etal,2018_Shimizu_etal,2018_Shimizu_etal_NJP,2018_Shimizu_etal_2}.

In particular, ultracold gases in optical lattices provide a versatile tool for simulating and studying dynamics of quantum many-body physics~\cite{1998_Jaksch_etal,2002_Greiner_etal,2008_Bloch_etal,2012_Cirac_Zoller,2012_Lewenstein_book},
including the early studies of single-component bosonic lattice gases~\cite{2018_Shimizu_etal,2018_Shimizu_etal_NJP,2018_Shimizu_etal_2}, Rydberg-dressed atoms~\cite{ZhouPhysRevA,Nature2019} and dipolar bosons~\cite{arXiv2021} in optical lattices.
Lots of interesting properties of these systems, such as verifying the Kibble-Zurek (KZ) scaling law between the defect formation and quench rate~\cite{2011_Chen_etal,2015_Braun_etal,2016_Anguez_etal,2016_Clark_etal}, have already been well studied. At the same time, there has also been a great interest in exploring the deviation from the KZ scaling in quench dynamics across the critical regime of the phase transition. Various new effects, such as new statistics of the resulting defects~\cite{2018_DelCampo,2020_GomezRuiz_etal}, fantastic evolution of correlation functions~\cite{2021_Roychowdhury_etal} and unusual saturation of the defect density~\cite{2016_Donadello_etal,2019_Ko_etal,2021_Goo_etal,2010_delCampo_etal,2018_Liu_etal,2015_Chesler_etal}
have been unveiled.

In this work, motivated by recent progresses in experimental investigation of the bosonic lattice gas inside a cavity~\cite{2016_Landig_etal,KlinderPhysRevLett}, where the effect of cavity-mediated interactions result in the observation of a rich equilibrium phase diagram including Mott insulator (MI), superfluid (SF), supersolid (SS) and charge-density-wave (CDW) phases, we study its non-equilibrium quench dynamics. This set-up offers new possibilities for exploring various non-equilibrium dynamics via varying the parameters of the Hamiltonian crossing different quantum phase transitions (QPT). And here we focus on studying the quench dynamics across the continuous phase transition between the CDW and SS phases. It is found that there is a threshold of the quench rate. Below that threshold, various physical
quantities, such as the correlation length and vortex density, show a good agreement with the KZ scaling law. While above that threshold, the saturation of the defect density has been found, indicating a deviation of the KZ scaling at fast quench.

This paper is organized as follows. In Section \uppercase\expandafter{\romannumeral2}, through employing the static Gutzwiller (GW) method, the equilibrium phase diagram is obtained, which is consistent with other mean-field calculations. {In Section \uppercase\expandafter{\romannumeral3}, the protocol of quench is introduced and  the non-equilibrium quench dynamics across the continuous phase transition between the CDW and SS phases have been studied by the time-dependent Gutzwiller (tGW) method}. The scaling laws of the correlation length and the number of defects have been investigated. It is shown that there is a threshold of the quench rate separating two distinct scaling regimes. One is satisfied with the KZ scaling law, the other is characterized with the saturation of defect density, showing a deviation from the KZ scaling. We further show that such distinct scaling behaviors can be understood through comparing the relaxation time and the quench rate.

{\section{Effective model and equilibrium phase diagram}}
Let us consider load a Bose-Einstein condensate (BEC) of $^{87}\mathrm{Rb}$ into a highly anisotropic $3$D optical lattice  coupled to an ultrahigh-finesse optical cavity, being similar as the ETH experimental setup~\cite{2016_Landig_etal}. Since the coherent scattering of light between the lattice and cavity modes creates a dynamical checkerboard superlattice for the atoms~\cite{2010_Baumann_etal,2012_Mottl_etal,2022_Wang_etal}, the effective Hamiltonian describing the atomic
dynamics dressed by the cavity field can be expressed as
\begin{eqnarray}
H&&=-J\sum\limits_{\left\langle {i,j} \right\rangle}\left( {\hat{b}_{{i}}^{\dag }{\hat{b}_{{j}}}+h.c.}\right) +\frac{U}{2}%
\sum\limits_{{i \in e,o}}{{{\hat{n}}_{i}}({{\hat{n}}_{i}}-1)}\nonumber \\
&&-\frac{U_{L}}{N}\left( \sum\limits_{{i \in e}}{{{\hat{n}}_{{i}}-}}%
\sum\limits_{{i \in o}}{{{\hat{n}}_{{i}}}}\right) ^{2}-\mu\sum\limits_{{i \in e,o}}{{{\hat{n}}_{i}}},%
\label{eq:Hamiltonian}
\end{eqnarray}
where $\hat{b}_{i}$ and $\hat{b}_{i}^{\dag}$ are the annihilation and creation operators for bosonic atoms at the lattice site ${\mathbf r}_{i}$. ${e}$ and ${o}$ refer to the even and odd sites of the lattice, respectively. $\hat{n}_{i}=\hat{b}_{i}^{\dag}\hat{b}_{i}$ is the on-site particle number operator. $N$ is the total number of lattice sites. $J$ captures the tunneling amplitude between nearest neighbors and $\mu$ is the chemical potential. The strength of the on-site repulsion is labeled as $U$, which can be tuned through the Feshbach resonance technics. $U_L=-N \hbar M_{0}^{2}\eta^{2}/\Delta_{c}$ describes the strength of the cavity-mediated interaction between atoms, where $M_{0}$ is the overlap between the Wannier function and the lattice potential $(\propto \cos\frac{2\pi x}{\lambda} \cos\frac{2\pi y}{\lambda})$ and $\eta$ is the two-photon Rabi frequency of the scattering process. $\Delta _{c}=\omega_{p}-\omega_{c}$ represents the dispersive shift of pumping frequency $\omega_{p}$ and cavity resonance frequency $\omega_{c}$.

The cavity-mediated interaction ($U_L$ term in Eq.~(\ref{eq:Hamiltonian})) favors an overall even-odd sites imbalance, which can be characterized by the defined charge-density-wave (CDW) order parameter as {$\theta=2\langle( \sum\limits_{{i \in e}}{{{\hat{n}}_{{i}}-}}\sum\limits_{{i \in o}}{{{\hat{n}}_{{i}}}})\rangle/N$}, where $\langle...\rangle$ means the expectation value in the ground state. The superfluid order parameters for the even and odd lattice sites are introduced as {$\phi_{e/o}=\left\langle b_{e/o}\right\rangle$}, respectively. Then, we employ the static Gutzwiller (GW) method to study the equilibrium zero-temperature phase diagram of the system. Consider starting with the GW ansatz
\begin{equation}
\left\vert {{\Psi _{gw}}}\right\rangle =\prod\limits_{i}{\sum\limits_{%
{n_{i}}=0}^{n_{max} }{f_{{n_{i}}}^{\left( {{i}}\right) }\left\vert {{n_{i}%
}}\right\rangle }},
\end{equation}
where $i\in e,o$ and $f_{n_{e}}^{(e)}$ , $f_{n_{o}}^{(o)}$ are the variational coefficients. $n_{max}$ is the maximum number of particles at each site. Then, the above defined order parameters can be rewritten as
\begin{eqnarray}
\phi _{i} &=& \underset{n_{i}=0}{\overset{n_{max}-1}{\sum }}\sqrt{n_{i}+1}f_{n_{i}}^{(i)\ast }f_{n_{i}+1}^{(i)}, \nonumber \\
{\theta} &=& {\frac {2}{N} \left( \sum\limits_{{i \in e}} \underset{n_{i}=0}{\overset{n_{max}}{\sum }}\left\vert f_{n_{i}}^{(i)}\right\vert ^{2}n_{i}-\sum\limits_{{i \in o}}\underset{n_{i}=0}{\overset{n_{max}}{\sum }}\left\vert f_{n_{i}}^{(i)}\right\vert ^{2}n_{i}\right)}.
\end{eqnarray}
{Here, we take $n_{max}=7$ for both even and odd sites, which is proved to be large enough for the study of our proposed system. The average number density of atoms is determined through the relation $\left\langle n_{i}\right\rangle = \underset{n_{i}=0}{\overset{n_{max}}{\sum }}\left\vert f_{n_{i}}^{(i)}\right\vert ^{2}n_{i}$. By minimizing the expectation ground state energy
$\langle {{\Psi _{gw}}}| H | {{\Psi _{gw}}}\rangle$, the variational coefficients in $|{{\Psi _{gw}}}\rangle$ can be determined (see Appendix A for details). Therefore, order parameters defined above can be obtained through the GW method. The equilibrium zero-temperature phase diagram can thus be obtained, such as shown in Fig. 1(a). The difference among distinct phases
can be captured by the order parameters, for example, as shown in Fig. 1(b). In the supersolid (SS) phase, a finite even-odd imbalance and nonzero superfluid order parameters coexist, which is distinct from a superfluid (SF) phase where the even-odd imbalance is vanished and
there is finite and equal superfluid order parameters. In the charge-density-wave (CDW) state, superfluid order parameters vanish. And the presence of a finite even-odd imbalance distinguishes the CDW from a Mott insulator
(MI) state. Our phase diagram is in good agreement with other existing studies~\cite{2016_Chen_etal,2016_Dogra_etal,2016_Niederle_etal,2016_Sundar_Mueller,2017_Flottat_etal,2019_Himbert_etal,2022_Wang_etal}.

\begin{figure}[tbh]
\includegraphics[width=8.6cm]{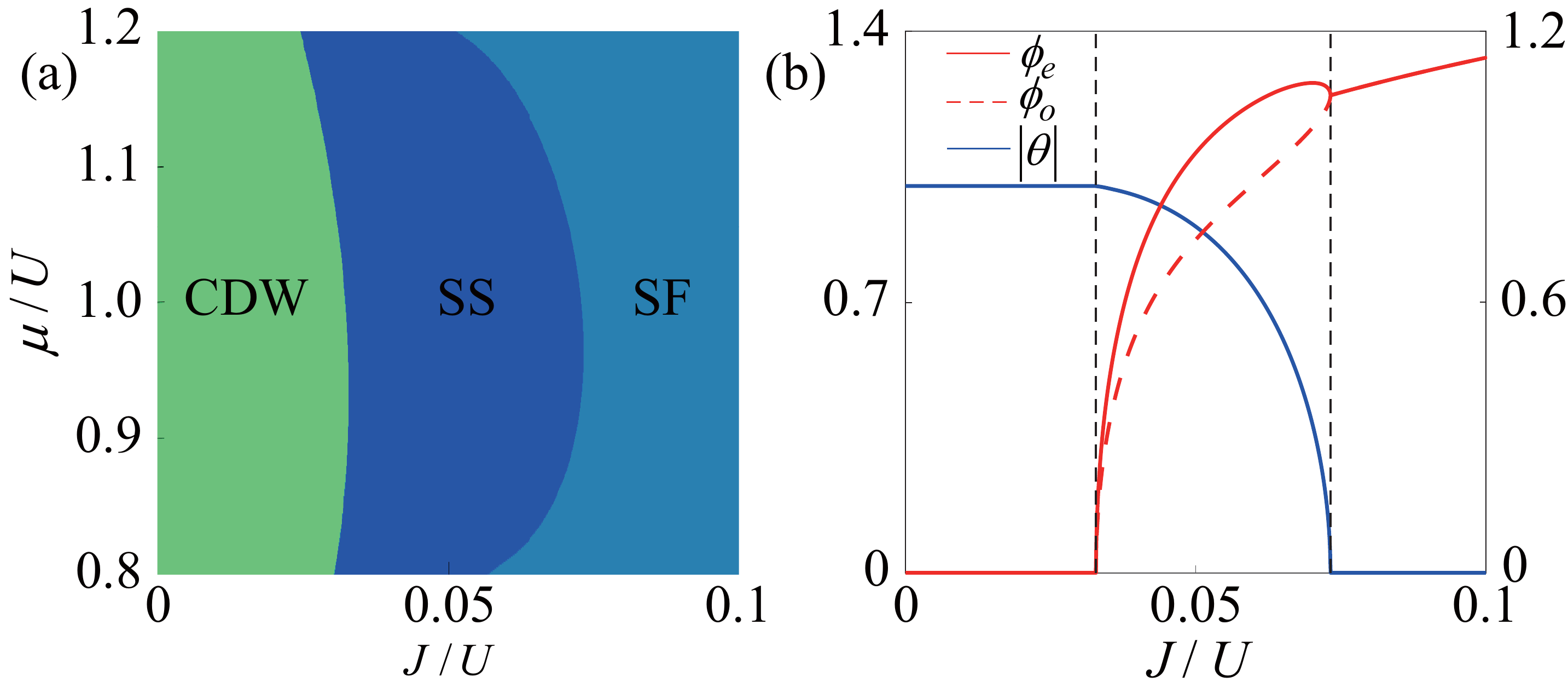}
\caption{\label{fig:wide} (a) Equilibrium
zero-temperature phase diagram as a function of the
hopping amplitude and chemical potential with $U_L/U=0.5$. (b) Superfluid order parameters $\phi_{e(o)}$ and the average imbalance $\theta$ between even and odd lattice sites as a function of hopping amplitude. Here we choose $\mu/U =0.95$ and other parameters are the same as in (a).}
\end{figure}

{\section{Quench dynamics across the phase transition from CDW to SS }}
In this section, we will study the dynamics across the phase transition from CDW to SS as shown in Fig. 1(a). The quench protocol is constructed through tuning the tunneling amplitude $J$ by the following relation
\begin{equation}
J(t)=\left\{
\begin{array}{c}
J_{i}+\frac{J_{f}-J_{i}}{2\tau _{Q}}(t+\tau _{Q})\text{ \ \ \ }t\in \lbrack
-\tau _{Q},\tau _{Q,}] \\
\text{ \ \ \ \ \ }J_{f}\text{ \ }\ \ \ \ \ \ \ \text{\ \ }\ \text{\ \ }t>\tau
_{Q}\label{eq:quench}\text{\ \ \ \ \ \ }%
\end{array}%
\right.
\end{equation}
where $J_i$ and $J_f$ are the initial and final values of the quench tunneling amplitude, which satisfy the relations $J_i<J_{c1}$ and $J_{c1}<J_f<J_{c2}$ with $J_{c1}$ and $J_{c2}$ being the equilibrium phase transition points of CDW to SS and SS to SF as shown in Fig. 1(a), respectively. Here $J_f+J_i=2J_{c1}$ means that the system approaches the quantum critical point  between CDW and SS
at $t =0$. $\tau _{Q}$ is the quench time, which captures the quench rate. We then employ the time-dependent Gutzwiller (tGW) method for investigating the real-time dynamics of our proposed system via varying $J(t)$ as in Eq.(4). In the practical calculation, we use the interaction strength $U$ as the
unit of energy and $\hbar/U$ is used as the  unit of time $t$. In the tGW approximation, the Hamiltonian in Eq. (1) is devised into a single-site Hamiltonian and can be approximated as $H_{GW}$ (see details in Appendix B). Then, we introduce the tGW wave function as
\begin{equation}
\left\vert {{\Psi _{tgw}}}\right\rangle =\prod\limits_{i}{\sum\limits_{%
{n_{i}}=0}^{n_{max} }{f_{{n_{i}}}^{\left( {{i}}\right)}(t)\left\vert {{n_{i}%
}}\right\rangle }}.
\end{equation}
The coefficients $f_{n_{i}}^{(i)}(t)$ can be determined by solving the following Schr$\ddot{o}$dinger equation
$i\hbar \partial _{t}\left\vert \Psi_{tgw} (t)\right\rangle =H_{GW}(t)\left\vert \Psi_{tgw}(t)\right\rangle$ for various initial states corresponding to the CDW phase and $H_{GW}(t)$ comes from replacing $J$ by $J(t)$ in $H_{GW}$. Therefore, the coefficients of tGW wave function can be determined through the following relation
\begin{eqnarray}
& \frac{\ i\hbar\partial {f_{{n}_{i}}^{({i})}(t)}}{{\partial t}}&
=-J(t)\sum\limits_{<{i,j}>}[{\phi }_{{j}}^{\ast }{\sqrt{{n}_{i}+1}f_{{n}_{i}+1}^{({i})}(t)+{\phi }_{{j}}%
\sqrt{{n}_{i}}f_{{n}_{i}-1}^{({i})}(t)}] \nonumber \\
&&+[{\frac{U}{2}{n}_{i}({n}_{i}-1)-\mu {n}_{i}}] {f_{{n}_{i}}^{({i})}(t)}-(-1)^{i}\theta{U_{L}}{n}_{i}f_{{n}_{i}}^{({i})}(t)\nonumber \\
\end{eqnarray}
Then, we use the fourth-order Runge-Kutta method to study the time evolution of the system
and calculate various physical quantities, such as the SF order parameter $\Phi$, correlation length ${\xi}$ and vortex number
$N_{v}$, which are defined as follows
\begin{eqnarray}
 &&\Phi =\frac{1}{N}\sum_{i} \left\vert \phi_{i} \right\vert, \notag \\
 &&\left\langle b^\dagger_i b_j \right\rangle \propto \exp\left(-\frac{|\mathbf{r}_i-\mathbf{r}_j|}{\xi}\right), \notag \\
 && N_{v}=\sum_{i}\left\vert \Omega _{i}\right\vert,
\end{eqnarray}
where $\Omega _{i}=\frac{1}{4}[\sin (\theta _{i+\hat{e}_{x}}-\theta _{i})+\sin
(\theta _{i+\hat{e}_{x}+\hat{e}_{y}}-\theta _{i+\hat{e}_{x}})+\sin (\theta _{i+\hat{e}_{y}}-\theta _{i+\hat{e}_{x}+\hat{e}_{y}})+\sin
(\theta _{i}-\theta _{i+\hat{e}_{y}})]$ with $\theta_{i}$ being the phase of superfluid order parameter $\phi_{i}$ and $\hat{e}_{x}(\hat{e}_{y})$ being the unit vector along the $x(y)$ direction. The system size is chosen as $100\times100$ in most of our numerical calculations and we have verified that this is a sufficiently large size where the size-dependent effect disappears.

\begin{figure}[tbh]
\includegraphics[width=7.6cm]{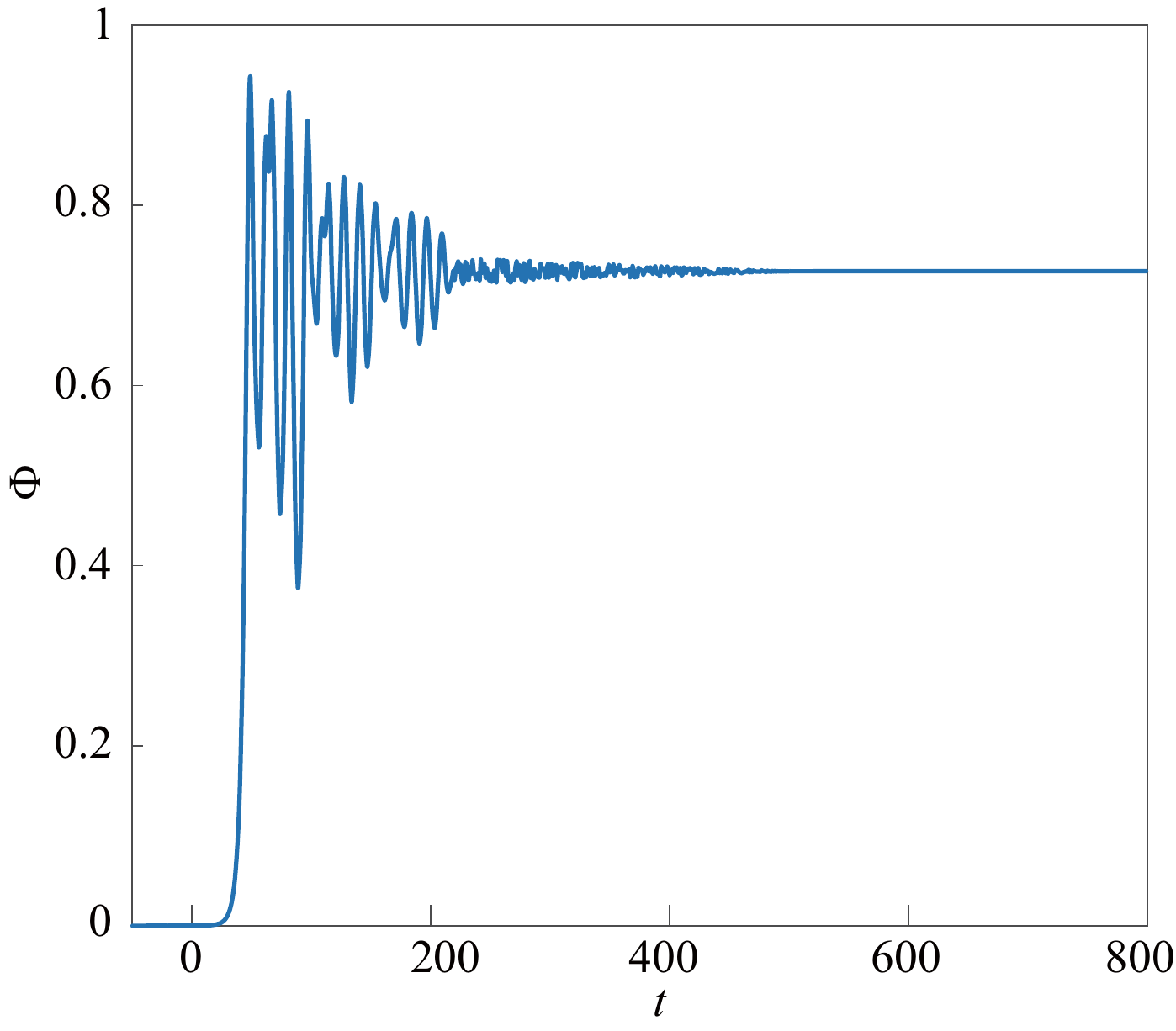}
\caption{\label{fig:wide} Time evolution of the defined SF order parameter $\Phi$. Here we consider
changing the hopping amplitude from $J_{i}/U=0$ to $J_{f}/U=0.0658$ linearly during the time interval $2\tau_{Q}$ with $\tau_{Q}=50$. Time is in the unit of $\hbar/U$ and
$\mu/U=0.95$. At $t= 0$, the system passes through the equilibrium phase transition point from CDW to SS.}
\end{figure}

As shown in Fig. 2, the typical behavior of SF order parameter $\Phi$ as the function of time $t$ is demonstrated. At $t=0$, the system crosses the quantum critical point between CDW and SS at $J_{c1}$ and the relaxation time of the system diverges. The dynamics is thus frozen and the state of the system cannot follow the change in the hopping amplitude. This scenario persists till the transition time $\hat{t}$, which is determined by the relation $|\Phi(\hat{t})|=2|\Phi({t=0})|$. Then, $\Phi$ develops very rapidly. After that rapid increase, $\Phi$ starts to fluctuate and such a oscillatory trend sets the average value tend towards the steady state value. We also study the defects formation during the course of the quench dynamics by investigating the vortex number $N_v$ defined above, as an indicator of the defect density. When the quench dynamics in the CDW region, the vortex density is high, which is caused by the effect of phase fluctuations introduced during the initial state preparation. Then, at $t=0$ the system crosses the quantum critical point and enters into the SS regime where spontaneous symmetry breaking occurs. The SF order parameters acquire finite values and small domains are gradually formed (shown in Fig. 3). However, such domains are formed where there is a phase coherence within, but not between the domains. As the quench is continued further, due to the annihilation of vortex-antivortex pairs, domains are merged and the size of domain becomes larger. The number of that thus decreases. After approaching the steady state, phase coherence is established in the system and the vortex number approaches zero.

\begin{figure}[tbh]
\includegraphics[width=8.9cm]{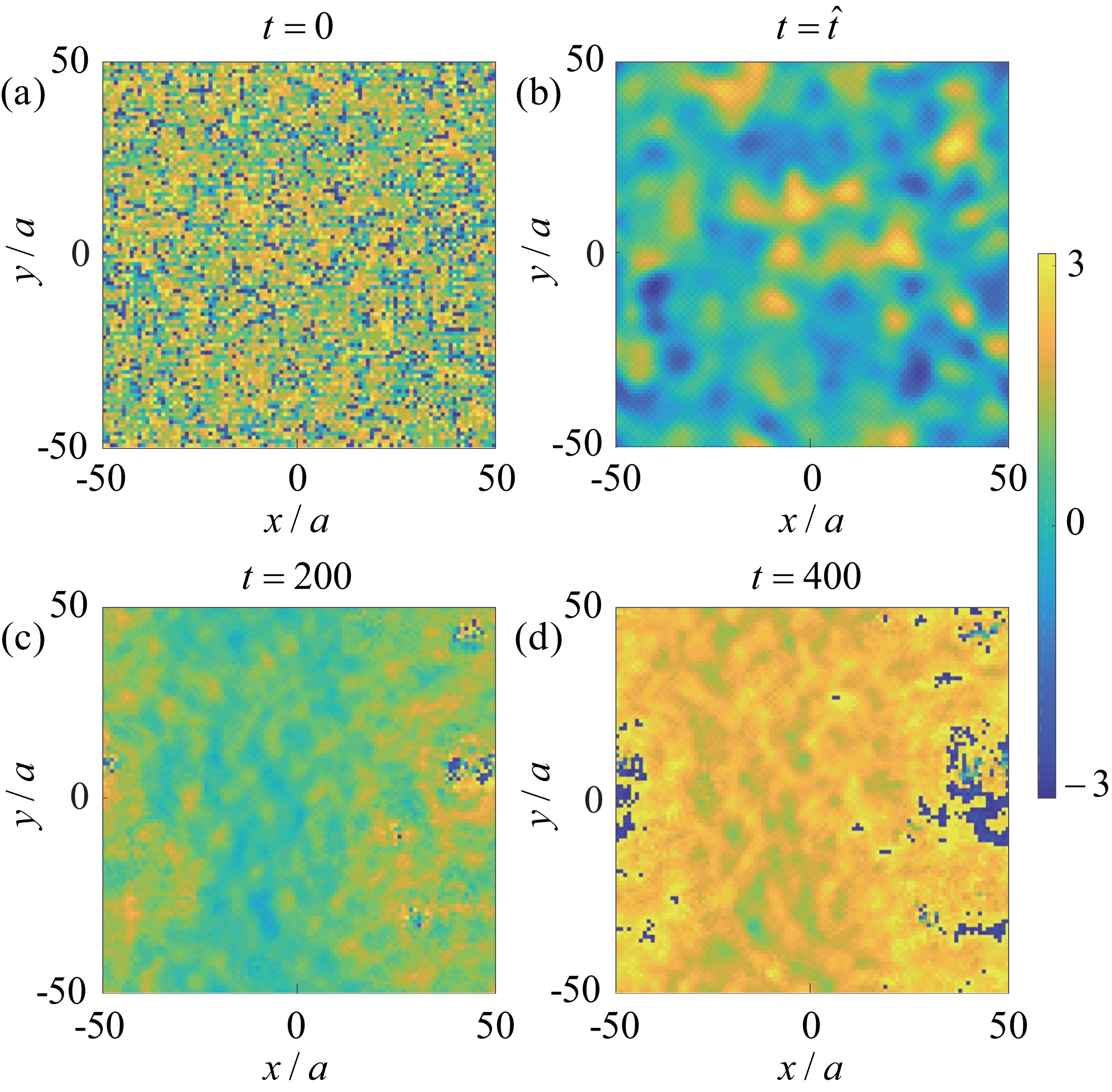}
\caption{\label{fig:wide} Snapshots of the phase of the order parameter $\phi _{i}$ at certain times
for $\tau{_Q}=50$. As shown in (a), when $t = 0$, the system crosses the quantum critical point and enters into the SS regime, where the SF order parameters acquire finite values and small domains is gradually formed. As shown in (b)-(d), when the quench is continued further, domains are merged. The size of that becomes larger and the number of that thus decreases. $a$ is the lattice constant. Other parameters are the same as in Fig. 2.}
\end{figure}

\begin{figure}[tbh]
\includegraphics[width=8cm]{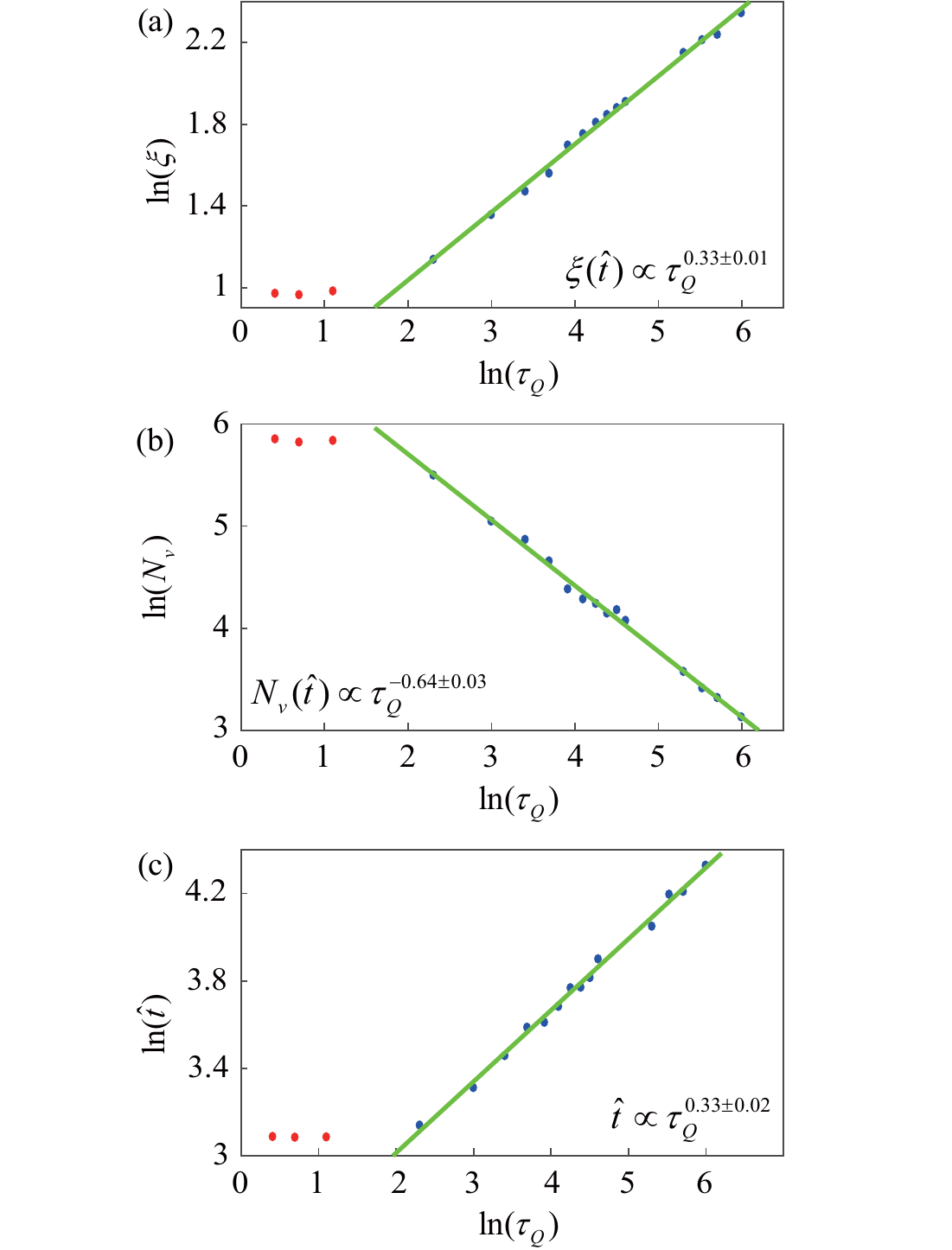}
\caption{\label{fig:wide} (a) and (b): Scaling law of the correlation length $\xi$ and vortex number $N_{v}$ at $t =\hat{t}$ with respect to $\tau_{Q}$. There are two distinct scaling regimes. One is in good agreement with the KZ scaling law, the other is characterized with the saturation of various quantities, indicating a deviation from the KZ scaling. (c) Scaling law of transition time $t =\hat{t}$  with respect to $\tau_{Q}$.
It has the similar behavior as (a) and (b). Other parameters are the same as in Fig. 2.}
\end{figure}

\begin{figure}[tbh]
\includegraphics[width=8.6cm]{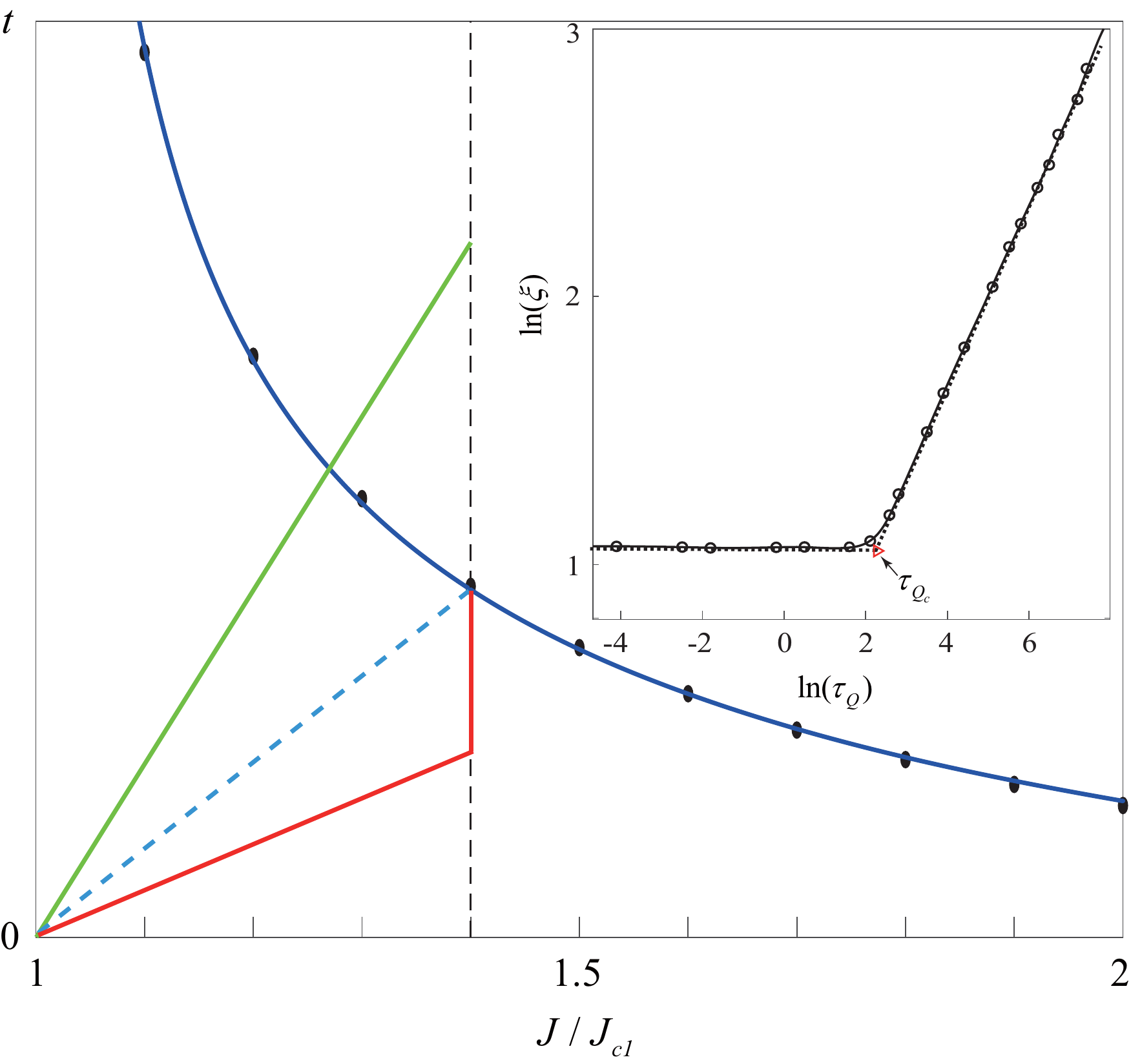}
\caption{\label{fig:wide} The relaxation time as a function of hopping amplitude (blue-solid line). The slope of blue-dashed line decides the threshold of quench rate separating two distinct scaling regime when fixing $J_f=\bar{J}$ with $\bar{J}/J_{c1}=1.4$. The red-solid line indicates one of the fast quench above the threshold, where a deviation from the
KZM predicted scaling occurs. The green-solid line stands for one of the slow quench below the threshold, where the adiabatic-impulse approximation is valid and the KZ scaling thus be satisfied. The inset shows that the intersection point of the un-smooth transition from the plateau to KZM power law scaling of the correlation length can also
determine the critical quench time $\tau_{Qc}$.}
\end{figure}

Next, we will investigate the behavior of quench dynamics with different quench rate $(\propto 1/\tau_Q)$. To study the scaling laws for the above defined quantities, such as correlation length ${\xi}$ and vortex number $N_v$, with respect to the quench time, we monitor ${\xi}$ and $N_v$ at the transition time $\hat{t}$, because $\hat{t}$ captures the freeze out time indicating the moment that the system catches the quench speed beginning to evolve quickly and can thus be linked to the universal scaling law connected with the relaxation time. As shown in Fig. 4, there are two different regimes separating distinct scaling behaviors for both ${\xi}$ and $N_v$ as a function of $\tau_Q$. For slow quench, such as $\tau_Q$ from $\tau_Q= 10$ to $\tau_Q= 400$ (Fig. 4), it is shown that both ${\xi}$ and $N_v$ satisfy a fairly good scaling law as ${\xi}\propto {\tau_Q}^{b}$ and ${N_v} \propto {\tau_Q}^{-d}$ with $b\approx0.33$ and $d\approx0.64$, respectively. The relation $d\approx2b$ holds. Therefore, the KZ hypothesis predicted power law scaling is satisfied in the slow quench regime. Moreover, the
KZM also predicts that the transition time $\hat{t}$ should have the scaling law as $\hat{t}\sim \tau_Q^{\nu z / (1+\nu z)}$ with $\nu$ and $z$ being the critical exponent of the equilibrium correlation length and the dynamical critical exponent, respectively.
Through extracting the exponent from the scaling of $\hat{t}$, such as shown in Fig. 4(c) and utilizing the relation $b=\frac{\nu}{1+\nu z}$ and $d=\frac{2\nu}{1+\nu z}$, $\nu$ and $z$ can be estimated. We get $z\approx0.98$ and $\nu \approx 0.49$, where $z$ is fairly close to that expected value $z=1$ from the 3D XY model, but $\nu$ dose not coincide with $\nu= 0.672$ predicated in the 3D XY model~\cite{PhysRevB2006}, indicating that the spatial inhomogeneity effects the critical behavior~\cite{2018_Shimizu_etal_2}.
While considering fast quench, as shown in Fig. 4, it is found that the vortex defect density
demonstrates a plateau whose value is a constant independent of the quench rate and ${\xi}(\hat{t})$ also has the similar behavior. It indicates that there is a violation of  the KZ scaling in the fast quench regime.

In the following, we will make a detailed analysis of the above distinct scaling behaviors and determine the critical quench rate ($\propto 1/\tau_{Qc}$) separating two different quench regimes. To understand the quench dynamics here, we first study the relaxation time of the system. Since the transition time $\hat{t}$ captures the moment that the system catches the quench speed and begin to evolve quickly, it can be linked to the relaxation time $\tau$ through the following relation
\begin{equation}
\hat{t}=\tau (J(\hat{t})).
\end{equation}
Therefore, through solving the above equation, the relaxation time can be obtained. As shown in Fig. 5, we display the relaxation time as a function of the hopping amplitude $J$. Through comparing the relaxation time and the quench rate, distinct scaling behaviors during the dynamic procedure can be understood. Let us consider fixing $J_f$ at a certain value at $J_f=\bar{J}$. Then, various quench procedures with different quench time $\tau_{Q}$ can be distinguished by a critical quench time $\tau_{Qc}$ determined by the relation $J(\hat{t})=\bar{J}$, for instance, as indicated by the blue-dashed line in Fig. 5. Therefore, $\tau_{Qc}$  can be determined by the slope of that blue-dashed line decides a threshold of the quench rate separating two distinct scaling regime. Let us first consider the fast quench above that threshold. For example, as indicated by the red-solid line in Fig. 5, the linear quench ends before the time that such line (stands
for $J(t)$) meets the $\tau(J)$ line. Therefore, the transition time $\hat{t}$ for any fast quench above the threshold of the quench rate will be the same and be fixed at $\hat{t} = \tau(\bar{J})$, being independent of the values of $\tau_{Q}$. The domains of defect for different quench rates will thus have the same average size, which naturally explains the plateau of defect density appears in rapidly quenched dynamics. While considering the slow quench below that threshold, for instance, as indicated by the green-solid line in Fig. 5, the $J(t)$ line intersects the $\tau(J)$ line before the end of linear quench. It means that the system cannot follow the change of $J(t)$ and the dynamics is frozen, which will persist till the transition time
marking the impulse period of the quench. After the evolution time is equal to the relaxation time, the system evolves promptly. Therefore, the adiabatic-impulse approximation is valid and the KZ scaling is satisfied. Furthermore, the critical quench time $\tau_{Qc}$ can also be determined by the intersection point of the un-smooth transition
from the plateau to KZM power law scaling of the correlation length or vortex density, for instance as shown by the dashed-line in the inset of Fig. 5. The obtained critical quench time $\tau_{Qc}$ is consistent with the way through comparing the relaxation time and the quench rate as mentioned above.

\vspace{14 pt}
\section{Conclusion}

In conclusion, we have examined the quench dynamics crossing a continuous phase transition between CDW and SS phases of a bosonic lattice gas with cavity-mediated interactions. By using the time-dependent Gutzwiller method, we find that various physical quantities, such as
the correlation length and vortex density, show two distinct scaling regimes with respect to the quench time. There is a threshold of quench rate. Below that, the scaling behavior is in good agreement with the KZ scaling law. When above that threshold, the scaling behavior is characterized with the saturation of various quantities, indicating a deviation from the KZ scaling. Furthermore, our proposal can
be viably simulated experimentally in ultracold gases, since such a system has already been implemented in experiments. It thus paves an alternative way for investigating the dynamics of the system combined with the cavity light and interacting quantum matters.

\begin{acknowledgments}
This work is supported by the National Key R$\&$D Program of China (2021YFA1401700), NSFC (Grant No. 12074305, 12147137, 11774282, 11950410491), the National Key Research and Development Program of China (2018YFA0307600), the Fundamental Research Funds for the Central Universities and Cyrus Tang Foundation Young Scholar Program (H. W., X. H., S. L. and B. L.). We also thank M. Arzamasovs for helpful discussions and the HPC platform of Xi'An Jiaotong University, where our numerical calculations was performed.
\end{acknowledgments}

\appendix

\section{Static Gutzwiller Method}
To employ the static Gutzwiller method for studying the equilibrium zero-temperature phase diagram of the system, we apply
the GW ansatz $\left\vert {{\Psi _{gw}}}\right\rangle$ to the model Hamiltonian in Eq.~\eqref{eq:Hamiltonian}. Then, the
expectation value for each terms in the Hamiltonian can be expressed in terms of $f_{n_{i}}^{(i)}$  by using the following
relations
\begin{widetext}
\begin{eqnarray}
\left\langle b_{i}^{\dag }b_{j}\right\rangle &=&( \underset{n_{i}=0}{%
\overset{n_{\max }-1}{\sum }}f_{n_{i}+1}^{(i)\ast }f_{n_{i}}^{(i)}\sqrt{%
n_{i}+1})(\underset{n_{j}=0}{\overset{n_{\max }-1}{\sum }}%
f_{n_{j}}^{(j)\ast }f_{n_{j}+1}^{(j)} \sqrt{n_{j}+1})\notag \\
\left\langle b_{i}^{\dag }b_{i}\right\rangle & =&\underset{n_{i}=0}{\overset{%
n_{\max }}{\sum }}\left\vert f_{n_{i}}^{(i)}\right\vert ^{2}n_{i}   \\
\left\langle b_{i}\right\rangle  &= &\underset{n_{i}=0}{\overset{%
n_{\max }-1}{\sum }}f_{n_{i}}^{(i)\ast }f_{n_{i}+1}^{(i)}\sqrt{n_{i}+1}%
   \notag
\end{eqnarray}
\end{widetext}
The expectation ground state energy $\langle {{\Psi _{gw}}}| H | {{\Psi _{gw}}}\rangle$ can thus be obtained as
\begin{widetext}
\begin{eqnarray}
\left\langle H \right\rangle &=&-J\underset{\left\langle i,j\right\rangle}{%
\sum}{\Big[}(\underset{n_{i}=0}{\overset{n_{\max }-1}{\sum }}%
f_{n_{i}+1}^{(i)\ast }f_{n_{i}}^{(i)}\sqrt{n_{i}+1}) (\underset{%
n_{j}=0}{\overset{n_{\max }-1}{\sum }}f_{n_{j}}^{(j)\ast }f_{n_{j}+1}^{(j)}%
\sqrt{n_{j}+1})
+(\underset{n_{j}=0}{\overset{n_{\max }-1}{\sum }}%
f_{n_{j}+1}^{(j)\ast }f_{n_{j}}^{(j)}\sqrt{n_{j}+1}) ( \underset{%
n_{i}=0}{\overset{n_{\max }-1}{\sum }}f_{n_{i}}^{(i)\ast }f_{n_{i}+1}^{(i)}%
\sqrt{n_{i}+1}){\Big ]} \notag \\
&-&\frac{U_{L}}{N}{\Big[}(\underset{i\in e}{{\sum }}%
\underset{n_{i}=0}{\overset{n_{\max }}{\sum }}\left\vert
f_{n_{i}}^{(i)}\right\vert ^{2}n_{i}) ^{2}+( \underset{i\in o}{%
{\sum }}\underset{n_{i}=0}{\overset{n_{\max }}{\sum }}%
\left\vert f_{n_{i}}^{(i)}\right\vert ^{2}n_{i}) ^{2}-2( \underset%
{i\in e}{{\sum }}\underset{n_{i}=0}{\overset{n_{\max }}{\sum }}%
\left\vert f_{n_{i}}^{(i)}\right\vert ^{2}n_{i}) ( \underset{i\in o}{%
{\sum }}\underset{n_{i}=0}{\overset{n_{\max }}{\sum }}%
\left\vert f_{n_{i}}^{(i)}\right\vert ^{2}n_{i}) {\Big ]}   \notag \\
&+&\frac{U}{2}\underset{i}{{\sum }}\underset{n_{i}=0}{%
\overset{n_{\max }}{\sum }}\left\vert f_{n_{i}}^{(i)}\right\vert
^{2}n_{i}(n_{i}-1) -\mu \underset{i}{{\sum }}\underset{n_{i}=0}{\overset{%
n_{\max }}{\sum }}\left\vert f_{n_{i}}^{(i)}\right\vert ^{2}n_{i}
\end{eqnarray}
\end{widetext}

By minimizing  $\langle {{\Psi _{gw}}}| H | {{\Psi _{gw}}}\rangle$  with respect to a set of amplitudes $f_{n_{i}}^{(i)}$,
we can determine the ground state of the system. Different order parameters defined in Eq. (3) can thus be calculated via $f_{n_{i}}^{(i)}$. The equilibrium zero-temperature phase diagram can thus be obtained as shown in Fig. 1(a).

\section{Time-dependent Gutzwiller method}
To study quench dynamics of the system, we employ the time-dependent Gutzwiller (tGW) methods. The tGW methods approximate
the Hamiltonian in Eq. (1) with a single-site Hamiltonian. To do that, we first treat with the cavity-mediated interaction term by the mean-field approximation as$-\frac{U_{L}}{N}\left( \sum\limits_{i \in e}{{{\hat{n}}_{i}-}}\sum\limits_{i \in o}{{{\hat{n}}_{i}}}\right)^{2}\approx\frac{U_LN \theta^2}{4}-U_{L}\theta\left( \sum\limits_{i \in e}{{{\hat{n}}_{i}-}}\sum\limits_{i \in o}{{{\hat{n}}_{i}}}\right)$, where $\theta$ describes imbalance between even and odd lattice sites as defined in Eq. (3). Then, the Hamiltonian in Eq. (1) can be approximated as
\begin{eqnarray}
H_{GW} &=&\underset{i}{\sum }H_{i} \notag \\
H_{i} &=&-J\underset{j\in iNN}{\sum }\left( \hat{b}_{i}^{\dag }\phi
_{j}+H.c.\right) +\frac{U}{2}%
\hat{n}_{i}(\hat{n}_{i}-1)-\mu \hat{n}_{i}\notag \\
&-&(-1)^{i}U_{L}\theta \hat{n}_{i},
\end{eqnarray}
{where $iNN$ denotes the nearest-neighbor($NN$) sites of $i$}. The dynamics of the system can be
determined by solving the following Schr$\ddot{o}$dinger equation
$i\hbar \partial _{t}\left\vert \Psi_{tgw} (t)\right\rangle =H_{GW}(t)\left\vert \Psi_{tgw}(t)\right\rangle$
via replacing $J$ by $J(t)$ in $H_{GW}$.

\bibliographystyle{apsrev}
\bibliography{EBHM_KZM}

%\nocite{*}

\end{document}